# Galileo in early modern Denmark, 1600-1650

Helge Kragh[*]

**Abstract**: The scientific revolution in the first half of the seventeenth century, pioneered by figures such as Harvey, Galileo, Gassendi, Kepler and Descartes, was disseminated to the northernmost countries in Europe with considerable delay. In this essay I examine how and when Galileo's new ideas in physics and astronomy became known in Denmark, and I compare the reception with the one in Sweden. It turns out that Galileo was almost exclusively known for his sensational use of the telescope to unravel the secrets of the heavens, meaning that he was predominantly seen as an astronomical innovator and advocate of the Copernican world system. Danish astronomy at the time was however based on Tycho Brahe's view of the universe and therefore hostile to Copernican and, by implication, Galilean cosmology. Although Galileo's telescope attracted much attention, it took about thirty years until a Danish astronomer actually used the instrument for observations. By the 1640s Galileo was generally admired for his astronomical discoveries, but no one in Denmark drew the consequence that the dogma of the central Earth, a fundamental feature of the Tychonian world picture, was therefore incorrect.

## 1. Introduction

In the early 1940s the Swedish scholar Henrik Sandblad (1912-1992), later a professor of history of science and ideas at the University of Gothenburg, published a series of works in which he examined in detail the reception of Copernicanism in Sweden [Sandblad 1943; Sandblad 1944-1945]. Apart from a later summary account [Sandblad 1972], this investigation was published in Swedish and hence not accessible to most readers outside Scandinavia. The same was the case with a careful study in which he examined how the pioneering ideas of Galileo Galilei (1563-1642) were received among Swedish astronomers and natural philosophers in the seventeenth century [Sandblad 1942]. Whereas in this note I call attention to the little known findings of Sandblad with regard to Galileo in Sweden, the main purpose is to cover, if only in a preliminary and incomplete way, some aspects of the corresponding Danish case. I look at how knowledge of Galileo and his telescopic discoveries was transmitted to and discussed by scholars in Denmark in the early period from 1610 to about 1650. I also call attention to the failed attempt of Tycho Brahe (1546-1601) to establish contact with Galileo a decade before he became a scientific celebrity.

---

[*] Centre for Science Studies, Department of Mathematics, Aarhus University, Aarhus, Denmark. E-mail: helge.kragh@ccs.au.dk.

It is worth recalling that at the time the kingdom of Denmark comprised also Norway, and Iceland as well, and that the Lutheran university in Copenhagen was the only one in Denmark-Norway. Although some of the professors and students were Norwegians, and a few were Icelanders, I have found no indication of an early interest in Galileo in the learned community in Norway itself [Dahl 2011]. It also needs to be pointed out that in the seventeenth century there was very little learned or scientific contact between scholars in Sweden and Denmark [Kragh et al. 2008; Danneskiold-Samsøe 2008]. The new ideas concerning physics and astronomy known collectively as "the scientific revolution" were transmitted to the two Scandinavian countries from the southern parts of Europe, primarily through letters and study travels. There was almost no direct scholarly contact between the two local centres of culture and science, the universities of Copenhagen and Uppsala.

## 2. First encounters: Bartholin and Granius

One of the most important and versatile Danish scholars in the early part of the seventeenth century, Caspar Bartholin (1585-1629) was a prolific writer of dissertations and textbooks. His main field of study was anatomy and his *Institutiones anatomicae*, first published 1611 and later in many reprints and new editions, earned him respect throughout Europe [Grell 1993]. However, he was equally at home in natural philosophy or what at the time counted as physics. In 1605-1611 he studied abroad, principally at the universities of Wittenberg, Leiden, Padua and Basel. From 1608 to 1610 he spent most of his time in Padua, mainly to study the medical sciences under the professors Hieronymus Fabricius (1533-1619) and Iulius Casserius (1552-1616). After having published *Institutiones anatomicae* in Basel, he returned to Copenhagen, where he was appointed professor of pedagogy. From 1613 to 1624 he served as professor of medicine, after which he advanced to the even more prestigious chair of theology.

While in Padua, Bartholin was most likely acquainted with Galileo, who since 1592 had been professor of mathematics at the university. However, the claim that he was a student of Galileo lacks documentation and is probably incorrect [Moesgaard 1972, p. 123]. In any case, in the spring of 1610 he witnessed the sensation caused by the publication of *Sidereus nuncius*, Galileo's famous book in which he reported his

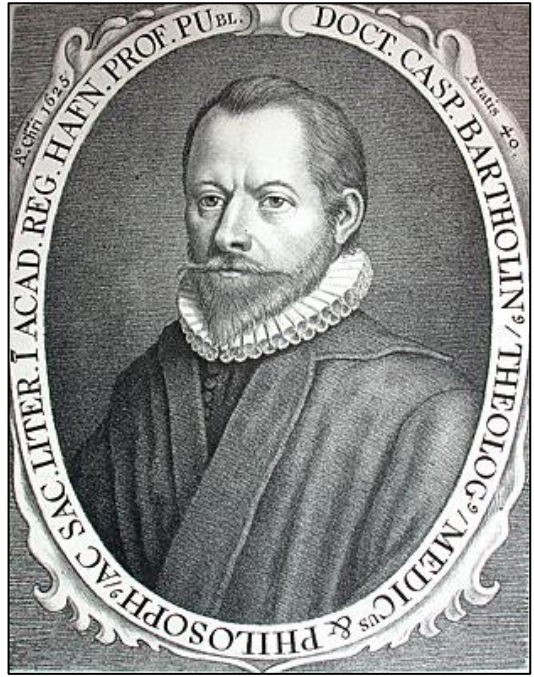
Fig. 1. Caspar Bartholin.

observations of the heavens with the new telescope. Not only did Bartholin know about the discoveries, he also had the opportunity to observe the night sky himself with the telescope, such as he mentioned in a treatise with the abbreviated title *De mundo* published in Rostock in 1617 [Bartholin 1617]. Eleven years later he incorporated the treatise together with other tracts on natural philosophy in the book *Systema physicum*, which for decades was widely used as a university textbook in Denmark.

Bartholin was impressed by some of the revelations of the telescope, but without accepting Galileo's arguments that they provided strong evidence in favour of the Copernican world system. Yet, despite his preference for Aristotelianism in natural philosophy he was open to new ideas as long as they could be brought to agree with Christian faith. For example, in his guide for students in medicine and physics, *De studio medico* of 1626, he mentioned three astronomical authorities of which Kepler was one. The other two were Tycho Brahe and his pupil Longomontanus, professor of astronomy in Copenhagen and Bartholin's brother-in-law. Given that Bartholin subscribed to the traditional Aristotelian-Ptolemaic world system it is surprising that he recommended the Copernican Kepler and the Tychonian Longomontanus, for in the latter's modified system the central Earth revolved around its axis. But Bartholin merely advised the students to learn these astronomical *hypotheses*, not to interpret them realistically.

4Although favourable to Tycho Brahe's astronomy and cosmology, Bartholin's observations with the telescope in 1610 convinced him that Tycho's understanding of the Milky Way was wrong while the one held by Galileo was right. In the words of Galileo, the Milky Way was "nothing but a congeries of innumerable stars distributed in clusters" [Galilei 1989, p. 62]. On the other hand, according to Tycho's view the Milky Way was an ethereal substance formed by residual primordial light, qualitatively similar to the stars but in a much more subtle and tenuous state that would not coagulate into denser bodies. Bartholin had formerly shared this view, but, as he said in *De mundo*, "thanks to having been instructed in Italy by the very ingenious Galilean telescope we have changed our mind; for when the eyes testify to questions concerning nature, the matter is made clear." Nonetheless, contrary to Galileo he did not regard the telescope to be cosmologically important and in his *Systema physicum* he launched an elaborate critique of the heliocentric hypothesis, in part based on physical arguments and in part on arguments relying on the Bible and the Mosaic tradition [Moesgaard 1972; Moesgaard 1977].

Caspar Bartholin's tract of 1617 was known to the Swedish scholar Æschillus Petræus (1593-1657), who became professor of theology in Uppsala and bishop in Åbo in Finland. In dissertations of 1625 he referred to and supported Bartholin's telescope-based understanding of the Milky Way. Petræus also seems to have been familiar with Galileo's *Sidereus nuncius*. Moreover, without mentioning Galileo by name, the telescope was briefly mentioned in a dissertation that Olaus Bureus (1578-1655) wrote in Basel in 1611 [Sandblad 1942, p. 117 and 121].

Whereas Bartholin may have been the first Scandinavian to look at the heavens with a telescope, he was not the only one to get acquainted with the new instrument in 1610. The Swede Nicolaus Andræ Granius (ca. 1580-1631) took his Magister degree at the University of Helmstedt in 1604, and in the autumn of 1610 he was visiting Prague. There he met the imperial astronomer Johannes Kepler (1571-1630), who just recently had put his hands on a telescope that Galileo had sent to the emperor Rudolf II (1552-1612). His observations with it confirmed Galileo's findings and caused Kepler to compose his *Dissertatio cum nuncio sidereo*, a work hastily written and published the same year. In an interesting letter of 3 October 1610 to his compatriot, the later arch bishop Johannes Lenæus (1573-1669), Granius writes about Galileo's discovery of the four "planets" revolving round Jupiter, adding that a telescope has recently arrived in Prague [Walde 1942]. Kepler had shown him the remarkable instrument which, he writes, enlarges distant objects by a factor of one thousand. It is not known what effect Granius' letter had, but it is hard to believe that Lenæus, who at the time was professor in Uppsala, did not share its content with his learned colleagues. In 1613 Granius was appointed professor of physics in Helmstedt, where he stayed until his death.

## 3. Two world systems: Longomontanus and Galileo

Danish astronomy in the first half of the seventeenth century was dominated by one of Tycho's most trusted assistants, Christian Sørensen (Christianus Severinus, 1562-1647) or better known as Longomontanus [Voelkel 2000; Christianson 2000, pp. 313-319]. After having joined Tycho and Kepler in Prague, Longomontanus returned to Denmark, where he was appointed professor of pedagogy and in 1607 professor of mathematics, a position that in 1621 was transformed into a new chair of astronomy (*mathematum superiorum*). Throughout his 40-year career as professor of mathematics and astronomy he endeavoured to defend and refine the Tychonian world system. His efforts were crowned with the publication in 1622 of *Astronomia Danica*, a book meant to be an alternative to Copernican as well as Ptolemaic astronomy. It was reprinted 1633 and 1640. Although generally faithful to Tycho's ideas, Longomontanus introduced in his system the non-Tychonian element of a spinning Earth as a substitute for the daily rotation of the heavens round the immobile Earth [Moesgaard 1972]. On the other hand, he firmly rejected the Copernican hypothesis of an annual motion of the Earth.

    In an appendix to *Astronomia Danica* on novel celestial phenomena Longomontanus mentioned several times Galileo and his telescopic observations, including his conclusions with regard to comets, new stars, the Milky Way, and mountains on the Moon [Longomontanus 1640, appendix, pp. 4-11]. Although this may have been the first time that Longomontanus referred in writing to Galileo, he undoubtedly was aware of him much earlier. He stayed with Tycho in Prague between January and August 1600 and thus presumably knew about Tycho's failed efforts to get in touch with Galileo and establish a correspondence with him. On 3 January 1600 Tycho wrote a letter to one of Galileo's friends, Giovanni Pinelli (1535-1601), urging him to contact Galileo on his behalf. Nothing happened and four months later Tycho addressed Galileo directly [Galilei 1929-1939, vol. 10, pp. 78-80]. At the time Galileo did not yet pursue astronomical research, but Tycho knew that he was a Copernican and may have thought that he did. At any rate, Galileo chose to ignore Tycho's approach [Heilbron 2010, pp. 117-119].

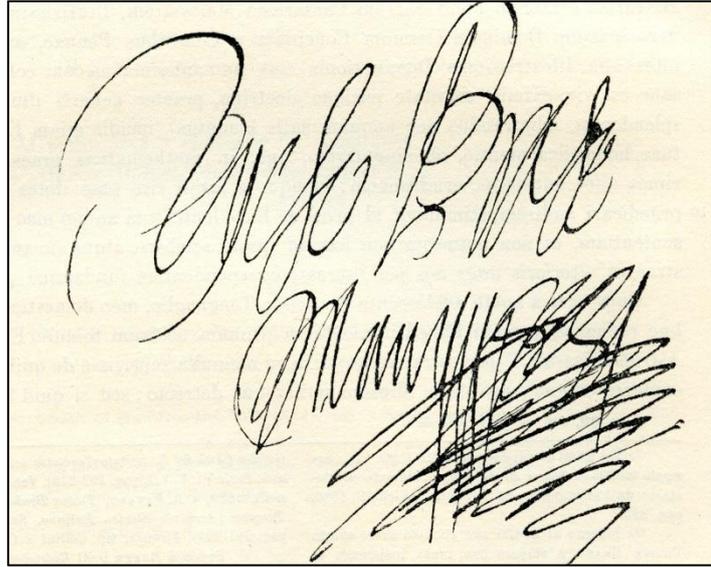

Fig. 2. Tycho Brahe's signature in his letter to Galileo
of 4 May 1600 [Galilei 1929-1939, vol. 10, p. 80].

After 1610, when Galileo engaged himself fully in astronomy and cosmology, he showed little direct interest in Tycho's system and none at all in Longomontanus' version of it. However, Tycho's name and ideas appeared prominently in two of his main works, *Il Saggiatori* from 1623 and the *Dialogo* from 1632, if mostly in a critical light. Moreover, he never mentioned explicitly the Tychonian world system by name. Nonetheless it was there, between the lines so to speak. What Galileo referred to as the non-Copernican system in *Dialogo* was, in many cases, in reality Tycho's rather than Ptolemy's system [Margolis 1991]. It seems clear that in general Galileo had little respect for the theories of the great Danish astronomer, but also that he wanted to avoid any confrontation with him [Heilbron 2010, pp. 247-250; Norlind 1970, p. 325]. The situation with regard to Longomontanus' system with a rotating Earth was different: although he was aware of it, he apparently considered it unimportant. He never mentioned it in his writings.

According to John Christianson's acclaimed book *On Tycho's Island*, as early as 21 June 1610 Longomontanus received a grant from the university "to build a telescope with some lenses" [Christianson 2000, p. 317]. If this was the case and Longomontanus actually built a telescope at this early time, it would be most remarkable. One might imagine that he was informed by Caspar Bartholin about Galileo's telescope and therefore wanted to construct one himself. Alas, this is pure speculation. Christianson's claim seems to rest on a misreading of the document describing a meeting among the Copenhagen professors on the mentioned date

7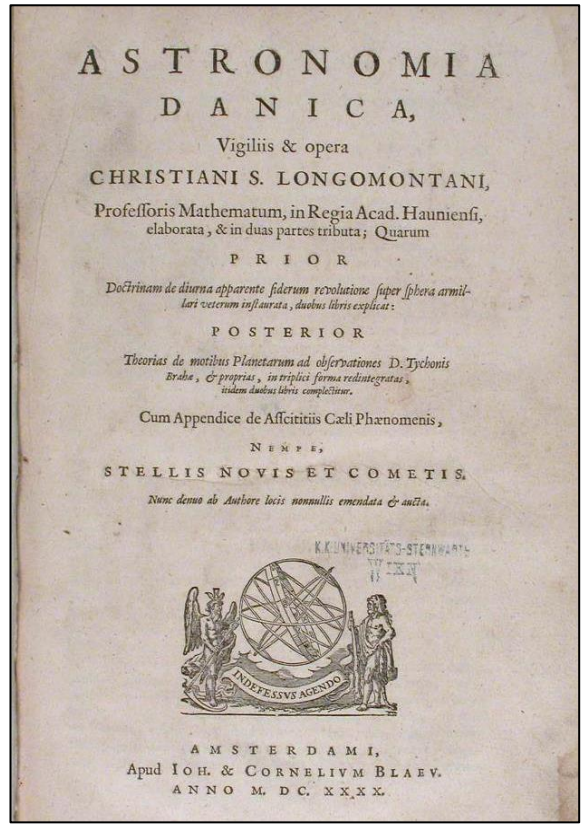

Fig. 3. The third edition of *Astronomia Danica*.

[Rørdam 1877, p. 367]. What Longomontanus applied for and received a small sum of money for was an unspecified *speculum* for use in his private residence. There is no mention in the document of either a telescope or optical lenses. In fact, there is no documentation for the claim that he ever possessed a telescope or made use of one for astronomical observations.

Longomontanus did discuss Galileo's telescope and its use in astronomy, but only in a tract from 1639 describing the purpose and design of the new Round Tower observatory that on the king's order was under construction in Copenhagen [Longomontanus 1639]. The tower was completed 1641 and ready for observations the following year. On the very last page of *Theatrum astronomicum* he added a section on the telescope, *De tubo optico*:

> This instrument, which is known as the telescope or, according to its form, the optical tube, has been … invented by brilliant Belgians; it was subsequently improved in a most excellent way by the Italian Galileo … and currently it has been produced in Naples and brought to Paris (as Morinus writes me) in such a size that it shows Mars as were it a Moon, and the Moon itself as seen from the Earth appears as a very large surface. We currently await results from this optical instrument that will surpass even

> those uncovered by Galileo's stellar telescope. … This optical apparatus … has not, according to my judgment, been very important with respect to the progress in astronomy. For astronomy does not so much investigate the heavenly bodies themselves and their casual properties as [it investigates] their motions and definite periods; it hands over the peculiarities of the stars to physics, which treats them by means of optics.

That is, even though Longomontanus admitted that the telescope offered some advantages, he did not consider it essential for astronomical purposes. It was of little use for positional astronomy and, contrary to the claim of Galileo, unable to provide decisive arguments in favour of either the Copernican world system or the rival systems of Ptolemy and Tycho Brahe. Morinus or Jean Baptiste Morin (1583-1656) was a famous but also controversial French mathematician, astronomer and astrologer with whom Longomontanus maintained a correspondence. Although strongly anti-Copernican he was also a correspondent of Galileo, whom he informed about Longomontanus' astronomical system.

Apart from his work in astronomy, Longomontanus was much occupied with mathematics and in particular with the classical problem of proving the quadrature of the circle. Indeed, he was obsessed with it. One of his early works on the subject, *Inventio quadraturae circuli* from 1634, was known to Galileo, although there is no reason to assume that he read it or was interested in Longomontanus' alleged proof [Galilei 1929-1939, vol. 16, p. 190]. In the introduction to one of his last works on the quadrature of the circle, *Rotundi in plano* published in 1644, Longomontanus stated that five years earlier he had written a letter to Galileo asking for his support [Kragh and Sørensen 2007, p. 28]. However, his attempt to win the support of Galileo failed. The 75-year old Galileo did not respond to the letter from the one year older astronomer in Copenhagen. He most likely never received it. Galileo was aware of Longomontanus and his *Astronomia Danica*, which appears in some of the letters reproduced in *Le Opere di Galileo Galilei* [Galilei 1929-1939, vol. 16, p. 190 and p. 252], but possibly for tactical reasons he ignored the revised Tychonian theory in his writings.

In the same year that Longomontanus published his *Astronomia Danica*, the astronomer Martinus Erici Gestrinius (1594-1648) at Uppsala University wrote the first Swedish dissertation that covered Galileo's discoveries. *De stellis* from 1622 dealt with topics such as the four moons encircling Jupiter, the rugged surface of the Moon, and the unusual tripartite appearance of Saturn in the telescope. In an astronomical textbook of 1647 called *Uraniae* Gestrinius dealt in more detail with Galileo's work, including a reference to the famous *Dialogo* from 1632 that led to the process against Galileo [Sandblad 1942; Nordenmark 1959, pp. 32-39]. *Dialogo* was of course banned, but Gestrinius used a Latin translation entitled *Systema cosmicum* that the German astronomer Matthias Bernegger (1582-1640) had produced in 1635, officially without Galileo's permission.

**4. Galileo in poetry**

Knowledge of the telescope and its new picture of the heavens permeated not only to mathematicians, physicists and astronomers, but also to the lay public that did not master Latin. Soon after Galileo announced his results in 1610 poets throughout Europe began praising them and incorporating elements of the new world picture in their art. In England, Galilei's discoveries entered poems by, among others, Andrew Marvell (1621-1678), John Milton (1608-1674) and John Donne (1572-1631), and poets in other countries followed suit [Nicolson 1956].

In Denmark we have an example in the minister and poet Anders Arrebo (1587-1637), whose *Hexaëmeron rhytmico-Danicum* was written around 1635 but only published in 1661. Although the title was in Latin, the poem itself was in Danish. Essentially an account of how God created the world in six days, the author also included astronomical and other scientific knowledge [Kragh et al. 2008, pp. 58-59]. Among the Danish authors he explicitly referred to were Tycho Brahe, Caspar Bartholin and Longomontanus. He also mentioned Copernicus' cosmological theory, if only to reject as "a phantasm of the brain" the absurd notion that the Earth rotated around its axis at great speed. Arrebo did not mention the Earth's annual journey round the Sun, perhaps because he considered it to be unimaginably absurd. Rendered into English, one of the verses in his *Hexaëmeron* reads:[1]

> No longer must we here aspire on Milky Way to tread
> to gaze upon *Galacti*'s seat with raised yet humble head;
> for that Way is but stars, so small in size but great in number,
> and so the Milky Path we need not venture to encumber.
> And if upon this truth you doubt, use Galileo's eye
> for thus you may yourself behold the twinkling stars on high.

This is probably the first reference in Danish language to Galileo's telescope. Arrebo was acquainted with Caspar Bartholin and his *Systema physicum*, from where he may have obtained his information. Other parts of the poem show that he had also read

---

[1] Translation by Heidi Flegal.

Bartholin's *De mundo* from 1617. A closer examination of Danish popular literature in the period may reveal more references to Galileo.

## 5. Frommius, Galileo and the telecope

A pupil of Longomontanus and eventually his successor as professor of astronomy, the young Dane Jørgen From or Georgius Frommius (1605-1651) prepared for an academic career by working as a tutor and guide for young students of wealthy parents travelling abroad [Kragh 2014]. In 1633, when Galileo was put under house arrest for life, Frommius was on his way from Padua to Paris. He undoubtedly heard about the infamous process against Galileo which created a stir throughout learned Europe, but we do not know how he reacted to the news. Some years later he was on another study travel that brought him and two students to the Netherlands, France and England. While in Leiden he reported the latest news about Galileo in a letter to the king's Chancellor Christen Friis (1581-1639):[2]

> Elzevir published recently Galileo's little book, which I dare to send to your Highness, since you asked me to keep my eyes open … . Galileo is still alive, but his health is poor. For more than a year he has blind on one of his eyes, and recently he has also lost the sight of the other. A mathematician from Amsterdam has been in Italy to speak with him about the determination of the longitude by means of Jupiter's moons. Galileo did not dare to write letters on this matter, and he is too old to travel abroad. It is inconceivable that many great men are persecuted, with the result that their most beautiful thoughts are thereby lost or remain unfinished. Consider for example Thomas Campanellus, who long ago was robbed of his right to think freely. And now Galileo is not allowed to publish his ideas in his own country. For this reason this treatise is published here without his knowledge, although in accordance with his wish, and it is distributed secretly, from one hand to another, by Elzevir junior.

Although Frommius, who subscribed to Longomontanus' version of the Tychonian world picture, disagreed with Galileo's cosmology, he defended his right to expressing his "beautiful thoughts" freely. The reference to "Campanellus" was to the Italian Dominican priest, philosopher and astrologer Tommaso Campanella (1568-1639), who spent the years 1599-1626 in the prison of the Neapolitan Inquisition. He nevertheless managed to write in 1622 an *Apologia pro Galileo* in which he defended Galileo's right to argue in favour of the heliocentric universe [Heilbron 2010, pp. 286-290; Campanella 1994]. However, the book was quickly banned and Campanella had to spend a couple of more years in the prison, this time in Rome. He finally obtained his freedom in 1629.

---

[2] Frommius' letters from his travel 1636-1638 are kept at the Royal Library in Copenhagen and have been translated by K. P. Moesgaard.

In another letter from the same period, addressed to the professor of theology Jesper Brockmand (1585-1652), Frommius confirmed that "the Elzevirs have printed a book by Galileo on mechanics [which] I will send to you." The Leiden publishing company Elsevier, established in 1580 by Ludwig Elsevier (Lodewijk Elzevier, 1540-1617), played a most important role in scientific and cultural circles in seventeenth-century Europe [Davies 1954]. Since Galileo could not publish his masterwork on mechanics *Discorsi intorno a due nuove scienze* in Italy, the manuscript was smuggled out to the Netherlands, where Elsevier published and distributed a Latin translation of it [Heilbron 2010, pp. 329-331]. The book was printed in Strasbourg and not in Leiden, as Frommius mistakenly thought. As we learn from his correspondence, at least two copies of Galileo's *Discorsi* were on their way to Denmark in 1638.

While in Leiden, Frommius practiced astronomy at the university's new observatory by means of a large quadrant and a smaller sextant. Of more interest is it that he also was in possession of or had access to a telescope. It did not belong to the observatory, which at the time was equipped only with traditional instruments. In a letter to Longomontanus of January 1638 he reported his observations with the telescope, probably the first such observations made by a Danish astronomer since Bartholin's more casual observation of 1610. He apparently continued his observations after having returned to Denmark, for in a work of 1642, *Dissertatio astronomica*, he writes: "I have often indulged in the practice of observing objects by means of the tube and I am in the possession of a tube of such a quality that objects are magnified almost hundred times, which provides a glimpse of all the things that were revealed to the celebrated Galileo regarding the planet Jupiter, spots on the Sun similar to those on the Moon, the various phases of Venus, and other things from current telescopic observations" [Frommius 1642]. While Frommius was happy to see with his own eyes what the great Galileo had announced many years ago, he also expressed scepticism with regard to astronomical claims based on the telescope alone. His attitude to the usefulness of the Galilean tube in astronomy did not differ significantly from Longomontanus'.

Frommius offered a more detailed discussion of the telescope in a follow-up work of 1645, *Responsio ad Johannis Baptistae Morini defensionem astronomiae restitutae*, and the subject was also covered in a dissertation of another young Danish scholar three years later. In *Exercitatio physico-mathematico* Christen Steenbuch (Christianus Steenbuchius, 1625-1665) discussed, much like Frommius had done earlier, the possible use of the telescope in astronomy. He kept his discussion in general terms and without any indication that he had used the instrument himself. Steenbuch mentioned the Neapolitan astronomer Francesco Fontana (ca. 1585-1656), who had improved Galileo's telescope but whose observations Galileo considered to be unreliable [Galilei 1929-1939, vol. 16, p. 18]. Moreover, the dissertation of 1648 described Galileo's discoveries of the Jovian moons and the spots on the Sun, mentioning also the rival discovery claims of Simon Marius (1573-1625) and Christoph Scheiner (1573-1650) and the resulting controversies with Galileo.

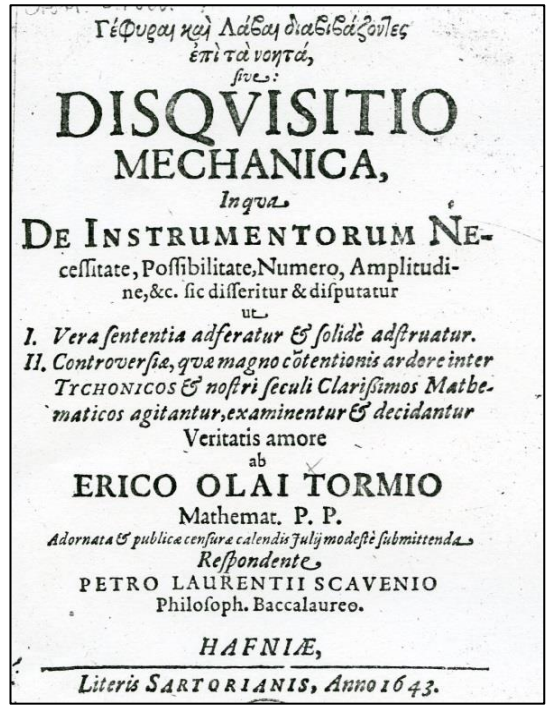

Fig. 4. One of the first Danish dissertations dealing with Galileo's astronomical ideas.

     The professor of mathematics Erik Olufsen Torm (Erico Olai Tormius, 1607-1667) wrote in 1643 a dissertation in which he defended the Tychonian system against "the most illustrious mathematicians of this century." He included among them Galileo and his arguments in *Dialogo de systemate mundi*, possibly the first reference to Galileo's famous book in Danish scientific literature.

     Nearly all Danish references to Galileo were to his telescope and astronomical discoveries. I have only come across a single reference alluding to other aspects of his work. While staying in Padua in 1642 as part of an extended study tour abroad, Thomas Bartholin (1616-1680), a son of Caspar Bartholin and soon a famous

anatomist and natural philosopher in his own right, wrote a letter to his uncle, the professor of medicine Ole Worm (1588-1654). Among the subjects of the letter was so-called Bononian stone [Kragh 2002]. The mysterious stone emitted a strange "cold light" – later identified as inorganic phosphorescence – that attracted much attention and caused much speculation. According to a book published by Fortunio Liceti (1577-1657), a professor from Bologna, the light was of the same kind as that emitted by the supposedly luminous Moon. He used the occasion to criticize Galileo's view that moonlight is just sunlight reflected from the Earth. Thomas Bartholin was acquainted with Liceti and in his letter to Worm he reported that Liceti had recently completed "his book against Galileo," which was published the same year as *De lunae subobscura luce prope coniunctiones*. Although Galileo never wrote a treatise on the Bononian stone, he was greatly interested in the phenomenon and got himself involved in a controversy with Liceti, whom he regarded as an exponent of the orthodox Aristotelian and hence unscientific world view [Drake 1978, pp. 405-412].

**6. Conclusion**

Whereas Galileo was well known and highly reputed in the first two decades of the seventeenth century, it took longer before he was discovered by astronomers and natural philosophers in the Nordic countries. Tycho Brahe was aware of him at an early date, but he was an exception. The first time Galileo was mentioned in print by a Danish scholar was in 1617, and five years later he appeared in a Swedish publication. Yet, still around 1640 there were only few references to his scientific work. What eventually attracted attention to the innovative Italian were almost exclusively his astronomical discoveries made by means of the amazing telescope. His advocacy of the Copernican world system was noted, but without making any impact. In the first half of the century there still were no Copernicans in either Denmark or Sweden. Astronomers were either Tychonians or supporters of the Ptolemaic theory.

     Galileo's international fame undoubtedly rested on his telescopic discoveries, but of course he also did pioneering work in mechanics and other branches of natural philosophy. First of all, he introduced the experimental method. There seems to be no mention in the Danish scholarly literature of the physical rather than astronomical Galileo. One looks in vain for awareness of or comments on his theory of the pendulum, his laws of freely falling bodies or his ideas about inertial motion; nor is his views on atomism, the void and the nature of heat to be found in the learned literature. These parts of Galileo's work were foreign to Danish natural philosophers

who predominantly thought in terms of Aristotelian concepts and tended to interpret the Bible quite literally. The situation in Sweden was not very different. Finally it is worth mentioning that apparently the process against Galileo in 1633 did not create much interest. It was known but not, as far as I can tell, discussed in print until much later.